\documentstyle[12pt,epsf]{article}

\def\lsim{\mathrel{\raise.2ex\hbox{$<$}\hskip-.8em\lower.9ex\hbox{$\sim$}}}
\def\gsim{\mathrel{\raise.2ex\hbox{$>$}\hskip-.8em\lower.9ex\hbox{$\sim$}}}
\let\alt=\lsim

\begin{document}

\thispagestyle{empty}

\renewcommand{\thefootnote}{\fnsymbol{footnote}}

\font\fortssbx=cmssbx10 scaled \magstep2
\hbox to \hsize{
\includegraphics{/NextLibrary/TeX/tex/inputs/uwlogo.ps}
\hskip.35in \raise.1in\hbox{\fortssbx University of Wisconsin - Madison}
\hfill$\vcenter{\hbox{\bf MAD/PH/958}
            \hbox{\bf astro-ph/9608185}
            \hbox{August 1996}}$ }

\vspace{.5in}

\begin{center} {\large\bf Active Galaxies as Particle Accelerators}\footnote{Talk given at {\it The VIIIth Rencontres de Blois, ``Neutrinos, Dark Matter and the Universe"}, Blois, France (1996), and at {\it Neutrino 96}, Helsinki, Finland (1996).}\\
\vspace{.15in}
Francis~Halzen\\
{\it Department of Physics, University of Wisconsin, Madison, WI 53706}\\
\end{center}

\vspace{.25in}

{\small\narrower
We present the theoretical arguments and describe the accumulating experimental evidence that jets, powered by supermassive black holes, are true cosmic accelerators. They produce photons of TeV energy, possible higher, and may be the enigmatic source of the highest energy cosmic rays. The features of the multi-wavelength emission spectrum are dictated by the interactions of electrons and protons, accelerated in the vicinity of the black hole, with the ambient light in the galaxy. Photoproduction of neutral pions by protons on UV light is the source of the highest energy photons, in which most of the bolometric luminosity of the galaxy may be emitted. They initiate an electromagnetic cascade which via pair production on the magnetic field and photon-photon interactions determines the emerging gamma-ray spectrum at lower energies. The lower energy photons, observed by conventional astronomical techniques, are, as a result of the cascade process, several generations removed from the primary high energy beams. The case that proton beams power active galaxies is far from conclusive. A three-prong technological assault on the problem will, in the near future, hopefully provide new insights into the structure of active galactic nuclei with the construction of second-generation atmospheric Cherenkov telescopes, large-scale EeV cosmic ray detectors and high energy neutrino telescopes. 
\par}

\vspace{.2in}

\def\large{\normalsize}
\def\Large{\normalsize}
\renewcommand{\thesection}{\arabic{section}.}

\section*{Introduction}

In recent years cosmic ray experiments have revealed the existence of cosmic particles with energies in excess of 10$^{20}$~eV. Incredibly, we have no clue where they come from and how they have been accelerated to this energy\cite{watson}. The highest energy cosmic rays are, almost certainly, of extra-galactic origin. Searching the sky beyond our galaxy, the central engines of active galactic nuclei (AGN) stand out as the most likely sites from which particles can be hurled at Earth with joules of energy. The idea is rather compelling because bright AGN are also the source of the highest energy photons, detected with air Cherenkov telescopes.

AGN are the brightest sources in the Universe; some are so far away that they are messengers from the earliest of times. Their engines must not only be powerful, but also extremely compact because their luminosities are observed to flare by over an order of magnitude over time periods as short as a day. Only sites in the vicinity of black holes which are a billion times more massive than our sun, will do. It is anticipated that beams accelerated near the black hole are dumped on the ambient matter in the active galaxy, mostly thermal photons with densities of 10$^{14}$/cm$^3$. The electromagnetic spectrum at all wavelengths, from radio waves to TeV gamma rays, is produced in the interactions of the accelerated particles with the magnetic fields and ambient photons in the galaxy. In early models the highest energy photons are produced by Compton scattering of accelerated electrons on thermal UV photons which are scattered up from 10~eV to TeV energy\cite{dermer}. The energetic gamma rays will subsequently lose energy by electron pair production in photon-photon interactions in the radiation field of the jet or the galactic disk. An electromagnetic cascade is thus initiated which, via pair production on the magnetic field and photon-photon interactions, determines the emerging gamma-ray spectrum at lower energies. The lower energy photons, observed by conventional astronomical techniques, are, as a result of the cascade process, several generations removed from the primary high energy beams.

High energy gamma-ray emission (MeV--GeV) has been observed from at least 40 active galaxies by the EGRET instrument on the Compton Gamma Ray Observatory\cite{EGRET}. Most, if not all, are ``blazars". They are AGN viewed from a position illuminated by the cone of a relativistic jet. Of the four TeV gamma-ray emitters identified by the air Cherenkov technique, two are extra-galactic and are also nearby blazars. The data therefore strongly suggests that the highest energy photons originate in jets beamed at the observer. A cartoon of an AGN, shown in Fig.~1, displays its most prominent features: an accretion disk of stars and gas falling into the spinning black hole as well as a pair of jets aligned with the rotation axis. Several of the sources observed by EGRET have shown variability, by a factor of 2 or so over a time scale of several days. Time variability is more spectacular at higher energies. On May 7, 1996 the Whipple telescope observed an increase of the TeV-emission from the blazar Markarian 421 by a factor 2 in 1 hour reaching eventually a value 50 times larger than the steady flux. At this point the telescope registered 6 times more photons from the Markarian blazar than from the Crab supernova remnant despite its larger distance ($10^5$) \cite{whipple}.

\begin{figure}[t]
\centering
\epsfxsize=3.25in\hspace{0in}\epsffile{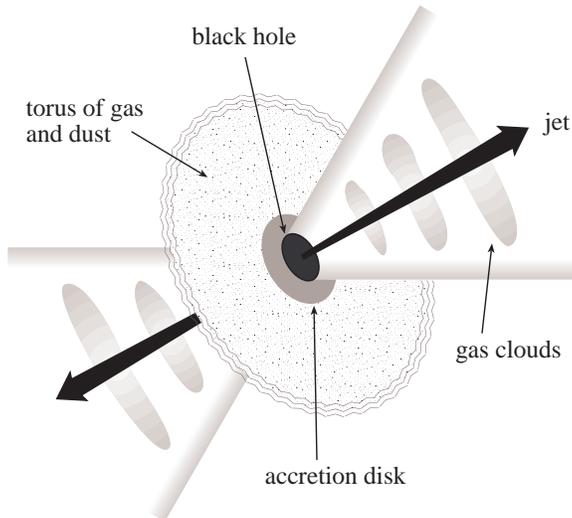}

\caption{Active galaxy with accretion disk and a pair of jets. The galaxy is powered by a central super-massive black hole ($\sim 10^9 M_\odot$). Particles, accelerated in shocks in the disk or the jets, interact with the high density of ambient photons ($\sim 10^{14}$/cm$^3$).}
\end{figure}

Before describing the models which attempt to accommodate this myriad of bewildering facts, one has to contemplate the effects of absorption. Energetic photons are efficiently decelerated by pair production of electrons on background light above a threshold
\begin{equation}
4E\epsilon > (2m_e)^2 \,,
\end{equation}
where $E$ and  $\epsilon$ are the energy of the accelerated and background photon, respectively\cite{footnote1}. Therefore TeV photons are absorbed on infrared light, PeV photons ($10^{15}$~eV) on the cosmic microwave background and EeV ($10^{18}$~eV) on radiowaves. It is likely that absorption effects explain why Markarian 421, at a distance of barely over 100~Mpc the closest blazar on the EGRET list, produces the most prominent TeV signal. Although the closest, it is one of the weakest; the reason that it is detected whereas other, more distant, but more powerful, AGN are not, must be that the TeV gamma rays suffer absorption in intergalactic space through the interaction with background infra-red light\cite{salomon}. This most likely provides the explanation why much more powerful quasars such as 3C279 at a redshift of 0.54 have not been identified as TeV sources. Many AGN may have significant very high energy components but that only Markarian 421 is close enough to be detectable with currently available gamma-ray telescopes. Undoubtedly part of the TeV flux is also absorbed on the infrared light in the source; we will return to this when discussing the blueprint of the accelerator.

Also protons interact with background light by the production of the $\Delta$-resonance just above the threshold for producing pions:
\begin{equation}
2E_p\epsilon > \left(m_\Delta^2 - m_p^2\right) \,.
\label{eq:threshold}
\end{equation}
The major source of energy loss of $\sim$100~EeV protons is photoproduction of the $\Delta$-resonance on cosmic microwave photons. The universe is therefore opaque to the highest energy cosmic rays, with an absorption length of only tens of megaparsecs when their energy exceeds $10^{20}$~eV. The photopion cross section grows very rapidly to reach a maximum of 540~$\mu$b at the $\Delta^+$ resonance $(s = 1.52$~GeV$^2)$. The $\Delta^+$ decays to $p \pi^0$ with probability of 2/3, and to $n\pi^+$ with probability 1/3. The neutral pions are a source of high energy photons, the charged pions of neutrinos. Lower energy protons, below threshold (\ref{eq:threshold}), do not suffer this fate. They cannot be used for astronomy however because their direction is randomized in the microgauss magnetic field of our galaxy.

Pion photoproduction may play a central role in blazar jets. If protons are accelerated along with electrons to PeV--EeV energy, they will produce high energy photons by photoproduction of neutral pions on the ubiquitous UV thermal background. Some have suggested that the accelerated protons initiate a cascade which also dictates the features of the spectrum at lower energy\cite{biermann}. From a theorist's point of view the proton blazar has attractive features. Protons, unlike electrons, efficiently transfer energy in the presence of the magnetic field in the jet. They provide a ``natural'' mechanism for the energy transfer from the central engine over distances as large as 1~parsec as well as for the observed heating of the dusty disk over distances of several hundred parsecs.

We will discuss the relative merits of the electron and proton blazar further on. Although these are a matter of intense debate, it is more relevant that the issues can be settled experimentally. The proton blazar is a source of high energy protons and neutrinos, not just gamma rays. Also, its high energy photon spectrum may exceed the TeV cutoff which is an unavoidable feature of the electron blazar. The opportunities for high energy neutrino astronomy are wonderfully obvious\cite{PR}.

Weakly interacting neutrinos can, unlike high energy gamma-rays and high energy cosmic rays, reach us from more distant and much more powerful AGN. Neutrino astronomers anticipate that the high energy neutrino sky will glow uniformly with bright active galaxies far outshining our Milky Way. The results may be even more spectacular. As is the case in man-made beam dumps, photons from celestial accelerators may be absorbed in the dump. The most spectacular sources may therefore have no counterpart in high energy photons. While neutrino astronomy is not too far in our future, existing ground-based gamma-ray and cosmic ray detectors are accumulating very suggestive evidence for the existence of proton blazars. We will return to this after a more detailed study of the accelerator.   

\section{Modelling of Blazar Jets}

Although no definite observations of photon emission beyond TeV energy have been reported, see however \cite{mannheim}, cosmic rays with much higher energies do exist and their origin is at present a complete mystery. The spectrum of the cosmic ray beam can be understood, up to perhaps 1000~TeV, in terms of acceleration by supernova shocks exploding into the interstellar medium. Although their spectrum suddenly steepens at 1000~TeV, a break usually referred  to as the ``knee", cosmic rays with much higher energies are observed and cannot be accounted for by this mechanism. This failure can be understood by simple dimensional analysis. We discuss this first.

Unlike the typical mono-energetic beam produced by a machine, cosmic accelerators produce power law spectra of high energy particles,
\begin{equation}
dN/dE\propto E^{-(\gamma+1)} \,. \label{eq:dN/dE}
\end{equation}
The observed high energy cosmic ray spectrum at Earth is characterized by $\gamma\sim 1.7$.  In general, a cosmic accelerator in which the dominant mechanism is first order
diffusive shock acceleration (first-order Fermi mechanism), will produce a spectrum with $\gamma \sim 1 +\epsilon$, where $\epsilon$ is a small number.  The observed cosmic ray spectrum is thought to be steeper than the accelerated spectrum because of the energy dependence of their diffusion in the galaxy: high energy ones more readily escape confinement. In highly relativistic shocks $\epsilon$ can take a negative value.

First-order Fermi acceleration at supernova blast shocks offers a very attractive model for a galactic cosmic accelerator, providing the right power and spectral shape. Acceleration takes time, however, because the energy gain occurs gradually as a particle near the shock scatters back and forth across the front gaining energy with each transit. The finite lifetime of the shock thus limits the maximum energy a particle can achieve at a particular supernova shock.  The acceleration rate is\cite{PR}
\begin{equation}
{\Delta E\over\Delta t} = K {v^2\over c} \, ZeB < ZeBc \,,
\label{eq:acc-rate}
\end{equation}
where $v$ is the shock velocity, $Ze$ the charge of the particle being accelerated and $B$
the ambient magnetic field.  The numerical constant $K\sim 0.1$ depends on the details of diffusion in the vicinity of the shock such as the efficiency by which power in the shock is converted into the actual acceleration of particles. The maximum energy reached is
\begin{equation}
E = {K\over c} (ZeB\,\ell v) < ZeB\,\ell \,.
\label{eq:Emax}
\end{equation}
The crucial time scale in converting Eq.~(\ref{eq:acc-rate}) into the limiting energy above is $\Delta t\sim v\,/\ell$, where $\Delta t \sim 1000$~yrs for the free expansion phase of a supernova. Using Eq.~(\ref{eq:Emax}) we ascertain that $E_{\rm max}$ can only reach energies of $\alt 10^{3}$\,TeV${}\times Z$ for a galactic field $B \sim 3\;\mu$Gauss, $K\sim 0.1$ and $v/c \sim 0.1$. Even ignoring all pre-factors the energy can never exceed $10^{17}$~eV by dimensional analysis. Cosmic rays with energy in excess of $10^{20}$~eV have been observed and the acceleration mechanism leaves a large gap of some five orders of magnitude that cannot be explained by the ``standard model'' of cosmic ray origin. To reach a higher energy one has to dramatically increase $B$ and/or~$\ell$. This argument is difficult to beat --- it is basically dimensional. Even the details do not matter; elementary electromagnetism is sufficient to identify the EMF of the accelerator or even the Lorentz force in the form of Eq.~(\ref{eq:Emax}).

The origin of cosmic rays with energy beyond the ``knee" is one of the oldest unresolved puzzles in science. Assuming that they are galactic, the measured spectrum implies that $10^{34}$ particles are accelerated to 1000~TeV energy and beyond every second. We do not know where or how. We do not know whether the particles are protons, iron nuclei or something else. If the cosmic accelerators indeed exploit the $3\;\mu$Gauss field of our galaxy, they must be much larger than supernova remnants in order to reach $10^{20}$~eV energy. Equation~(\ref{eq:Emax}) requires that their size be of order 30~kpc which exceeds the dimensions of our galaxy. Although imaginative arguments exist to avoid this impasse, it is generally believed that our galaxy is too small and the magnetic fields too weak to accelerate the highest energy cosmic rays. The gyroradii of the highest energy particles observed exceed the size of our galaxy; they should therefore point back at their sources. As there is no experimental evidence that their arrival directions are correlated to the plane of the galaxy, an attractive alternative is to look for large size accelerators outside the galaxy. Nearby active galactic nuclei distant by order 100~Mpc are the obvious candidates.

We already described features of blazar jets identifying them as true cosmic accelerators. One can visualize the accelerator in a very economical way in the Blanford-Zralek dynamo model. Imagine that the horizon of the central black hole acts as a rotating conductor immersed in an external magnetic field. By simple dimensional analysis this creates a voltage drop
\begin{equation}
{\Delta V\over 10^{20}{\rm volts}} = {a\over M_{\rm BH}} \; {B\over 10^4{\rm G}} \; {M_{\rm BH}\over 10^9 M_\odot} \,,
\end{equation}
corresponding to a luminosity
\begin{equation}
{{\cal L}\over10^{45}\rm erg\,s^{-1}} = \left(a\over M_{\rm BH}\right)^2
\left(B\over 10^4\rm G\right)^2 \left(M_{\rm BH}\over 10^9M_\odot\right)^2\,.
\end{equation}
Here $a$ is the angular momentum per unit mass taken to be the black hole mass $M_{\rm BH}$. Whether conversion of the rotational energy of the spinning black hole is responsible for the observed power of the AGN or not, the argument illustrates how it is dimensionally possible that particles may be accelerated to energies characteristic of the most energetic cosmic rays. The example hardly solves the problem however because it neglects efficiency factors likely to be much smaller than unity; see Eq.~(\ref{eq:acc-rate}). Particles carrying joules of energy remain somewhat of a mystery.

\section{The Multi-Wavelength Spectrum of Blazars}

Confronted with the challenge to explain a relatively flat multi-wavelength photon  emission spectrum reaching TeV energies which are radiated in bursts of a duration of days, models have converged on the blazar blueprint shown in Fig.~2. Particles are accelerated by Fermi shocks in blobs of matter travelling along the jet with a bulk Lorentz factor of order $\gamma \sim 10$. In order to accommodate bursts lasting a day in the observer's frame, the size of the blob must be of order $\gamma c \Delta t \sim 10^{-2}$~parsecs. The blobs are actually more like sheets, thinner than the jet's size of roughly 1~parsec. The $B$-field is taken to be of order 10~Gauss. The observed radiation at all wavelengths is produced by the interaction of the accelerated particles in the blob and the ambient radiation in the AGN which has a significant component concentrated in the so-called ``UV-bump\rlap".

\begin{figure}[t]
\centering
\epsfxsize=3in\hspace{0in}\epsffile{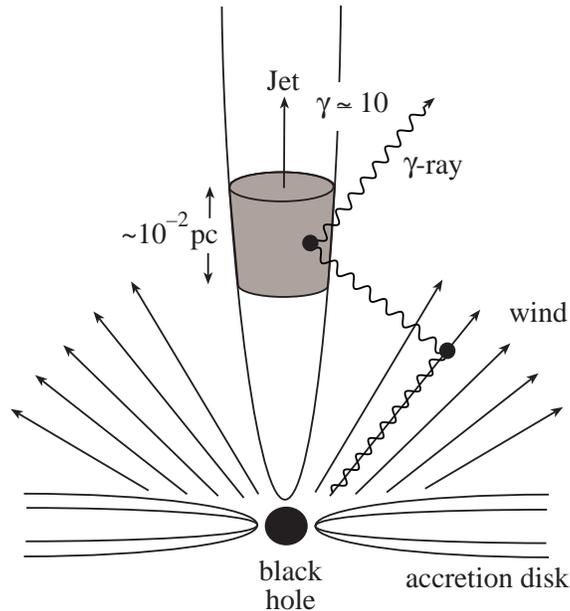}

\caption{Possible blueprint for the production of high energy photons and neutrinos near the super-massive black hole powering an AGN. Particles accelerated in sheets or blobs moving along the jet interact with photons radiated by the accretion disk or, produced by the interaction of the accelerated particles with the magnetic field of the jet.}
\end{figure}

The characteristic photon emission spectrum as well as the UV thermal radiation are shown in Fig.~3. The data is for the blazar 3C279. The luminosity is displayed in such a way that a flat spectrum corresponds to $\gamma\sim 1$, characteristic of acceleration of the electron beam by shocks. Notice that most of the bolometric luminosity is emitted at the highest energy. Three components in the photon spectrum are clearly visible and consist of synchrotron radiation produced by the electron beam on the $B$-field in the jet, synchrotron photons Compton scattered to high energy by the electron beam and, finally, UV photons Compton scattered by the electron beam to produce the highest energy photons in the spectrum\cite{dermer}. The seed photon field can be either external, e.g.\ radiated off the accretion disk as in the previous example, or result from the synchrotron radiation of the electrons in the jet, so-called synchrotron-self-Compton models.

\begin{figure}[t]
\centering
\epsfxsize=4.5in\hspace{0in}\epsffile{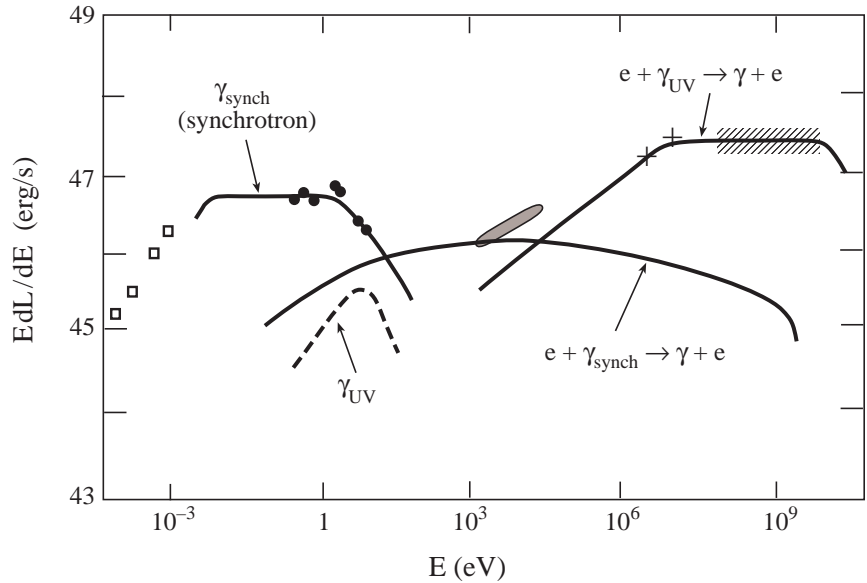}

\caption[]{Inverse Compton blazar from reference\cite{dermer}. Ambient UV light and synchrotron photons are scattered to high energy by Compton scattering of electrons accelerated along the jet on ambient light. The synchrotron photons are produced by interaction of the same electron beam with the magnetic field of the jet. The multi-wavelength flux is for the active galaxy 3C279, see reference\cite{dermer} for details.}
\end{figure}

Although the model describes the features of the observed spectrum as can be seen in Fig.~3, the picture has a variety of problems. In order to reproduce the experimental high energy luminosity the accelerating blobs have to be positioned very close to the black hole. The photon target density is otherwise insufficient for inverse Compton scattering to produce the observed flux. This is a balancing act however because the same dense target will efficiently absorb the high energy photons by $\gamma\gamma$ collisions. The balance is difficult to arrange especially in light of observations showing that the high energy photon flux extends beyond TeV energy\cite{whipple}. The natural cutoff occurs in the 10--100~GeV region; see Fig.~3. Finally, in order to prevent the electrons from losing too much energy before producing the high energy photons, the magnetic field in the jet has to be artificially adjusted to less than 10\% of what is expected from equipartition with the radiation density.

\section{The Proton Blazar}

For these and the more general reasons already mentioned in the introduction, the proton blazar has been developed. In this model protons as well as electrons are accelerated and because of their reduced energy loss protons can produce the high energy radiation further from the black hole. The more favorable production-absorption balance far from the black hole makes it relatively easy to extend the high energy photon spectrum above 10 TeV energy, even with bulk Lorentz factors that are significantly smaller than in the inverse Compton models. Two recent incarnations of the proton blazar illustrate that these models can also describe the multi-wavelength spectrum of the AGN\cite{mannheim, protheroe}; see Fig.~4. Because the seed density of photons is still much higher than that of target protons, the high energy cascade is initiated by the photoproduction of neutral pions by accelerated protons on ambient light via the $\Delta$ resonance; see Eq.~(\ref{eq:threshold}). The protons collide either with synchrotron photons produced by electrons\cite{mannheim}, or with the photons radiated off the accretion disk\cite{protheroe} as shown in Fig.~2.

\begin{figure}[h]
\centering
\epsfxsize=4in\hspace{0in}\epsffile{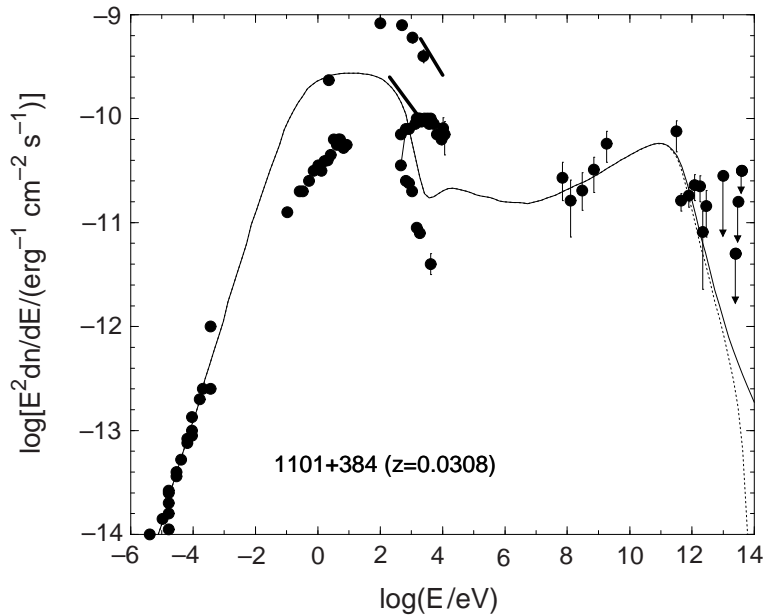}

\caption[]{The multi-wavelength spectrum of  Markarian 421 is compared with the proton blazar model of  Mannheim. The model accommodates the TeV observations of Whipple and HEGRA, see reference\cite{mannheim} for details.}
\end{figure}

Accelerated protons reach energies of $10^3$~PeV or more: With a $B$-field of 10~Gauss and a size of $10^{-2}$~parsecs, a factor $10^4$ larger and smaller than a supernova blastwave respectively, energies reach $10^{18}$~eV. We here used Eq.~(\ref{eq:Emax}) and took into account a Lorentz factor $\gamma = 10$. In this model the energy is actually restricted by the limited time over which the protons can be accelerated before losing energy on the ambient light:
\begin{equation}
{1\over \Delta t} \propto E_p B^2 \,,
\label{eq:timelimit}
\end{equation}
which states that the acceleration time is inversely proportional to the density of the photon field seen by the protons. In this model equipartition between energy density and $B$-field can be maintained; therefore the photon density is proportional to $B^2$. It is also assumed that the target photon spectrum is described by a power law with $\gamma = 1$. The number of photons above the photoproduction threshold of Eq.~(\ref{eq:threshold}) therefore grows linearly when the proton energy $E_p$ is increased; hence the first factor in Eq.~(\ref{eq:timelimit}). Combining Eqs.~(\ref{eq:acc-rate}) and (\ref{eq:timelimit}) we obtain
\begin{equation}
E_p \lsim {10^{10}{\rm~GeV}\over \sqrt {B(\rm Gauss)}} \,.
\label{eq:E_p}
\end{equation}
For $B \sim 10$~Gauss we again obtain proton energies as large as $10^3$~PeV. Do not forget that all energies have to be boosted by a Lorentz factor $\gamma$ relative to an external observer.

\begin{figure}[h]
\centering
\epsfxsize=4in\hspace{0in}\epsffile{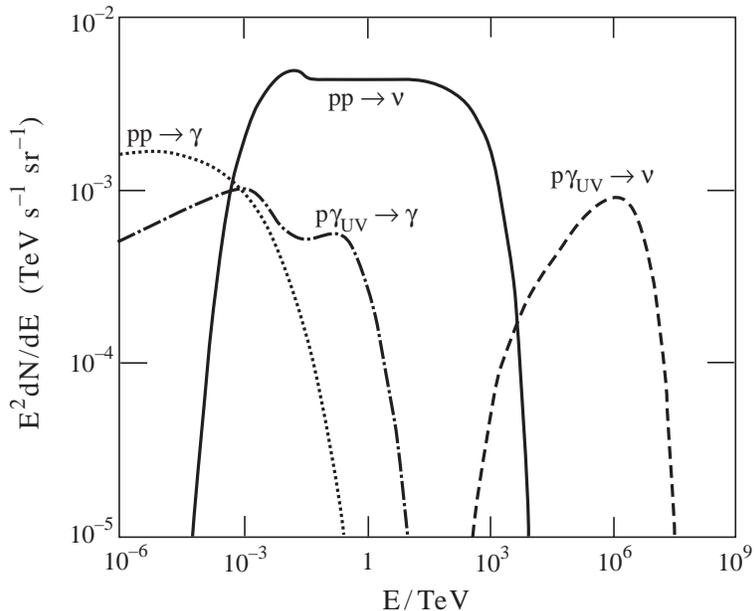}

\caption[]{Neutrino fluxes predicted by the proton blazar model of reference\cite{protheroe}. Shown are the photon and neutrino fluxes from pion photoproduction by accelerated protons on UV light. Also shown are predictions for the production of photons and neutrinos on ambient protons, these are however very uncertain. The curves are normalized such that in each case $E^2dN/dE = 1$ at 1~GeV for photons.}
\end{figure}

In the proton blazar radiation and magnetic field are in equipartition, the maximum energy matches the $B\ell$ value expected from dimensional analysis and, finally, the size of the blob is similar to the gyroradius of the highest energy protons. There is no fine-tuning. Predictions for gamma and neutrino emission spectra are shown in Fig.~4 and 5. In Fig.~4 the predictions of the proton blazar are confronted with the data on Markarian 421. In Fig.~5 the anticipated neutrino fluxes are shown with the warning that the estimates for production on ambient nucleons are very uncertain. Worth noting are the increase of the gamma-ray cutoff into the TeV energy range as well as the ``burst shape" of the neutrino spectrum with an average energy of
\begin{equation}
\left< E_\nu \right> \simeq {1\over20} (\gamma E_p) \simeq 10^9\rm\ GeV\,,
\end{equation}
using Eq.~(\ref{eq:E_p}) and $B = 10$~Gauss. Normalizing the flux to the observed EGRET sources yields about 1 neutrino of $10^3$~PeV energy per day in a kilometer-scale detector, neglecting neutrinos produced on the protons in the jet\cite{hill}. Neutrino telescope builders should take note that although there are less neutrinos produced than in the generic AGN models of a few years ago\cite{stecker}, they are all above PeV energy where the detection efficiency is increased and the atmospheric neutrino background negligible. If confirmed, these models strongly favor the construction of neutrino telescopes following the distributed architecture with large spacings of the optical modules\cite{halzen}.

\section{Evidence for the Proton Blazar?}

Discovery that active galaxies are proton accelerators will radically transform astronomy and astrophysics in one of its most fundamental areas of research. There are 3 obvious opportunities for corroborating the evidence which is at present, admittedly, inconclusive: i) correlate the direction of the highest energy cosmic rays to AGN, ii)~detect gamma rays with energy in excess of $\sim 10$~TeV and, finally, iii) detect neutrinos. With the rapidly expanding Baikal and AMANDA detectors producing their first hints of neutrino candidates\cite{domogatsky, hulth}, observation of neutrinos from AGN would establish the production of pions and identify the acceleration of protons as the origin of the highest energy photons. For the first two options, the future may be now.

Astronomy with protons becomes possible once their energy has reached a value where their gyroradius in the microgauss galactic field exceeds the dimensions of the galaxy. Protons with $10^{20}$~eV energy point at their sources with degree-accuracy. At this energy their mean-free-path in the cosmic microwave background is unfortunately reduced to only tens of megaparsecs. A clear window of opportunity emerges: Are the directions of the cosmic rays with energy in excess of  $\sim 5 \times 10^{19}$~eV correlated to the nearest AGN (red-shift $z$ less than 0.02), which are known to be clustered in the so-called ``super-galactic" plane? Although far from conclusive, there is some evidence that such a correlation exists\cite{stanev}. Lack of statistics at the highest energies is a major problem. Future large aperture cosmic ray detectors such as the new Utah HIRES air fluorescence detector and the Auger giant air shower array will soon remedy this aspect of the problem\cite{watson}.

We have already drawn attention to the 10~TeV photon energy as the demarkation line between the electron and proton blazars. The 10~GeV cutoff in the inverse Compton model, evident in Fig.~3, can be pushed to the TeV range to accommodate the Whipple data on Markarian 421, but not beyond. Bringing the accelerator closer to the black hole may yield photons in excess of 10~TeV energy, they have, however, no chance of escaping without energy loss on the dense infrared background at the acceleration site. HEGRA has been monitoring the 10 closest blazars, including Markarian 421, with its dual telescopes: the scintillator and naked photomultiplier detector arrays. The announcement\cite{rhode} that their upper limit for the aggregate emission of photons of 50~TeV energy and above may be a signal, may provide compelling evidence that blazar jets are proton accelerators; see Fig.~6.

\begin{figure}[t]
\centering
\epsfxsize=4in\hspace{0in}\epsffile{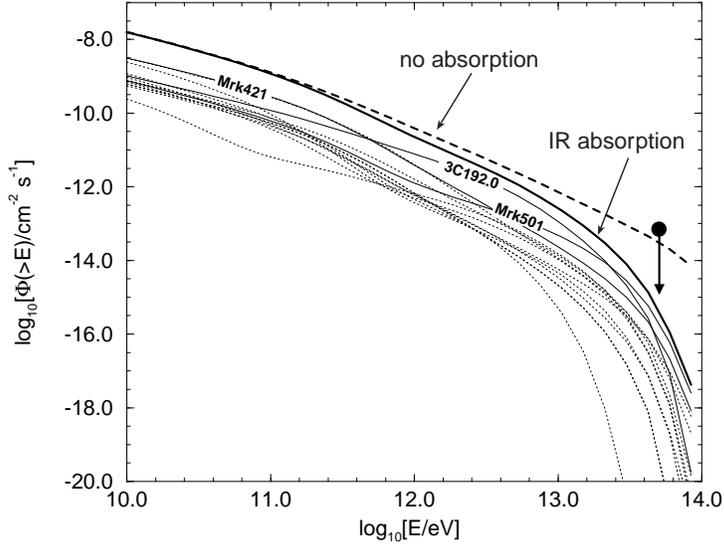}

\caption[]{Total photon flux from the 10 closest AGN is shown before and after taking into account the absorption of the TeV flux on interstellar infrared light\cite{mannheim}.}
\end{figure}

Model-independent evidence for cosmic proton accelerators can be obtained by establishing that they are sources of high energy neutrinos. A relatively model-independent estimate of the required telescope area can be made by computing the expected neutrino rate from the assumption that the observed EeV cosmic rays are produced by AGN. Because of the strict limitations on the density of target photons previously discussed, it is easy to show that roughly one neutrino is produced for every accelerated proton in the beam. This balance is easy to understand once one realizes that in astrophysical beam dumps the accelerator and production target form a symbiotic system. Although larger target mass may produce more neutrinos, it also decelerates the protons producing them. This is yet another aspect of the delicate acceleration-absorption balance. Equal neutrino and proton luminosities are therefore typical for the astrophysical beam dumps considered\cite{PR} and imply that
\begin{equation}
4\pi \int dE (E \, dN_\nu/dE) \sim L_{\rm CR} \sim 10^{-9}\rm\ TeV\ 
cm^{-2}\ s^{-1} \;.
\end{equation}
The luminosity $L_{\rm CR}$ has been conservatively estimated from cosmic ray data by assuming that only the highest-energy component of the cosmic-ray flux above 10$^{17}$~eV is of AGN origin. These particles, with energies beyond the ``ankle" in the spectrum, are almost certainly extra-galactic and are observed with a $E^{-2.71}$ power spectrum. Assuming an $E^{-2}$ neutrino spectrum, the equality of cosmic-ray and neutrino luminosities implies:
\begin{equation}
E{dN_\nu\over dE} = {1\over4\pi} {10^{-10}\over E\,(\rm TeV)} \, \rm 
cm^{-2}\ s^{-1}\ sr^{-1} \;. \label{eq:flux}
\end{equation}
Roughly the same result is obtained by assuming equal numbers of neutrinos and protons rather than equal luminosities. It is clear that our estimate is rather conservative because the proton flux reaching Earth has not been corrected for absorption of protons in ambient matter in the source and in the interstellar medium.

The probability to detect a TeV neutrino is roughly $10^{-6}$\cite{PR}. It is easily computed from the requirement that, in order to be detected, the neutrino has to interact nearer than the range of the muon it produces. In order for the neutrino to be detected, the muon has to reach the detector. Therefore,
\begin{equation}
P_{\nu\to\mu} \simeq {R_\mu\over \lambda_{\rm int}} \simeq 10^{-6} E_{\rm TeV}^2 \,,
\end{equation}
where $R_{\mu}$ is the muon range and $\lambda_{\rm int}$ the neutrino interaction length. The above equation is valid only in the energy range where both the range and cross section depend linearly on energy. Combining the 2 last equations we obtain that neutrino detector with 10$^6$~m$^2$ effective area is required for observing 100 up-coming muons per year, or maybe 10 from a nearby source. This estimate agrees with the earlier one based on the detailed proton blazar model. A kilometer-scale neutrino detector may be required\cite{halzen}.

In summary, the evidence is mounting that active galaxies may be true particle accelerators with proton beams dictating the features of the spectrum. They may be the sources of the highest energy cosmic rays. A topic which is one of the most fundamental in astronomy may have solutions driven by particle physics. A three-prong technological assault on the problem will hopefully lead to deeper insights in the near future with the construction of next-generation air Cherenkov telescopes, the Auger giant air shower array and deployment of the first stages of large neutrino detectors in natural water and ice.

\section*{Acknowledgements}

This talk benefitted from discussion with Peter Biermann, Manuel Drees, Charles Goebel, Karl Mannheim, Raymond Protheroe and Todor Stanev.
This work was supported in part by the University of Wisconsin
Research Committee with funds granted by the Wisconsin Alumni Research
Foundation, and in part by the U.S.~Department of Energy under Grant
No.~DE-FG02-95ER40896.


\newpage
\begin{thebibliography}{99}
\let\sl=\it

\bibitem{watson}
The Pierre Auger Project Design Report, Fermilab report (1995) and references therein.

\bibitem{dermer}
M.~Sikora, M.~C.~Begelman and M.~J.~Rees, {\it Ap. J. Lett.} {\bf 421}, 153 (1994) and references therein.

\bibitem{EGRET}
D.~J.~Thompson et al., {\it Ap. J. S}, {\bf 101}, 259.

\bibitem{whipple}
M.~Punch {\em et al.}, {\sl Nature} {\bf 358}, 477--478 (1992), D.~J.~Macomb et al., {\it Ap. J.} {\bf 438}, 59; {\bf 446}, 99 (1995).

\bibitem{footnote1}
The angular dependence of the collision cross section has been neglected.

\bibitem{salomon}
F.~W.~Stecker, O.~C.~De Jager and M.~H.~Salamon,
{\it Ap.\ J.} {\bf 390}, L49 (1992).

\bibitem{biermann}
K.~ Mannheim and P.~L.~ Biermann, {\it Astron. Astrophys.} {\bf 221}, 211 (1989)

\bibitem{PR}
T.~K.~Gaisser, F.~Halzen and T.~Stanev, {\it Physics Reports} {\bf 258}, 173 (1995).

\bibitem{mannheim}
K.~Mannheim, S.~Westerhoff, H.~Meyer and H.~H.~Fink, Astronomy and Astrophysics, in press (1996) and references therein.

\bibitem{protheroe}
R.~J.~Protheroe, Gamma Rays and Neutrinos from AGN Jets, Adelaide preprint ADP-AT-96-4 (1996).

\bibitem{hill}
G.~C.~Hill, {\it Astroparticle Physics}, submitted (1996)

\bibitem{stecker}
F.W.~Stecker, C.~Done, M.H.~Salamon and P.~Sommers, {\it Phys.\ Rev.\ Lett.} {\bf 66}, 2697 (1991) and {\bf 69}, 2738(E) (1992).

\bibitem{halzen}
F.~Halzen, The Case for a Kilometer-Scale Neutrino Detector: 1996, in {\it Proc.\ of the Sixth International Symposium on Neutrino Telescopes}, ed.\ by  M.~Baldo-Ceolin, Venice (1996).

\bibitem{domogatsky}
G.~V.~Domogatsky, in {\it Proc.\ of the Sixth International Symposium on Neutrino Telescopes}, ed.\ by  M.~Baldo-Ceolin, Venice (1996).
 
\bibitem{hulth}
P.~O.~Hulth for the AMANDA collaboration, Neutrino 96, Helsinki (1996).

\bibitem{stanev}
T.~Stanev et al.,  {\it Phys. Rev. Letters} {\bf 75}, 3056 (1995)

\bibitem{rhode}
W.~Rhode, High Energy Neutrino Workshop, Aspen, Colorado (1996)


\end{thebibliography}
\end{document}